\documentstyle[11pt,newpasp,twoside,epsf]{article}
\def\edcomment#1{\iffalse\marginpar{\raggedright\sl#1\/}\else\relax\fi}
\reversemarginpar

\begin{document}
\title{Non-LTE line-formation for CNO}
\author{Norbert~Przybilla, Rolf-Peter~Kudritzki}
\affil{Institute for Astronomy,\,2680 Woodlawn Drive,\,Honolulu,\,HI\,96822,\,USA}
\author{Keith~Butler, Sylvia~R.~Becker}
\affil{Institut f\"ur Astronomie und Astrophysik, Universit\"atssternwarte, 
Scheinerstr. 1, D-81679 M\"unchen, Germany}

\begin{abstract}
Accurate atomic data have become available 
in the recent
past due to the demands of astrophysics and fusion research. We report on
the impact of such data on non-LTE line-formation calculations for CNO in
early-type stars. Considerable improvement is achieved by the derivation of
consistent results from practically all available spectroscopic indicators,
regardless of ionization stage or spin system, and the uncertainties in the
analyses are drastically reduced. Moreover, systematic
trends are revealed, e.g.~an increase of the N\,{\sc i} abundances from 
previous studies of BA-type supergiants by a factor of two is indicated. The 
present work promises stringent observational constraints on 
chemical mixing in the course of massive star evolution. First results
on BA-type supergiants in the Galaxy and the Magellanic Clouds are discussed.
\end{abstract}

\section{Introduction}
Carbon, nitrogen and oxygen represent the body of
the heavy elements in the universe. Knowledge of their abundances is a 
key ingredient for understanding the evolution of stars, galaxies 
and the universe. However, reliable and accurate information on CNO
abundances is scarce. In the late-type stars it is primarily the 
convective nature of the atmospheres that introduces systematic 
uncertainty, such that even the solar CNO abundances are under debate still 
(e.g.~M.~Asplund, these proceedings). On the other hand, non-LTE effects 
prevail in the earlier spectral types. Besides efficient numerical
techniques, accurate atomic data are required for solving the non-LTE 
radiative transfer problem. Vast amounts of radiative and collisional data 
from quantum-mechanical {\em ab-initio} calculations have been provided by 
the Opacity Project (Seaton et al.~1994) and the IRON Project (Hummer et
al.~1993) in the recent years, supplemented by contributions from fusion
research and state-of-the-art experiments. Use of the $R$-matrix method in the
close-coupling approximation allows for a reduction of the typical 
uncertainties in the data to $\sim$10\% on the mean, in sharp contrast to the
order-of-magnitude approximations widely used in astrophysics.
In the following we discuss the impact of such data on non-LTE abundance
determinations for CNO in early-type stars and report on first applications.

\section{Non-LTE line-formation for CNO: recent progress}  
Comprehensive model atoms for C\,{\sc i/ii}, N\,{\sc i/ii} and O\,{\sc i} 
are implemented (Przybilla, Butler, \& Kudritz\-ki~2001; Przybilla \& 
Butler~2001; Przybilla et al.~2000; Przybilla~2002) for the non-LTE 
line-formation programs {\sc Detail} and {\sc Surface} (Giddings~1981, with 
substantial modifications by K.~Butler). For the first time also detailed 
excitation cross-sections for electron collisions are accounted for in 
hundreds of transitions, besides accurate radiative data, adopted from 
sophisticated quantum-mechanical computations.

Testing of the models is performed on high-resolution and high-S/N Echelle 
spectra of several bright galactic main sequence stars and supergiants of 
spectral types B and A, using line-blanketed {\sc Atlas9} model atmospheres 
(Kurucz~1993). Similar non-LTE effects are at work in all three elements.
Photoionizations depopulate the low-lying energy levels of the neutral 
species, while recombination cascades provide a significant overpopulation 
of the (quasi-)metastable states in the line-formation region. Moreover, 
photon losses -- in particular in the tenuous atmospheres of the 
supergiants -- lead to a drop of the line source function below the 
Planckian value. Both effects promote the non-LTE strengthening of the CNO
lines of the neutral atoms at visual and near-IR wavelengths. 
Non-LTE abundances will therefore be systematically smaller than derived 
in LTE. Non-LTE strengthening is also found for the
lines of the singly-ionized species.

The new results confirm these well-known facts from 
previous studies. However, for the first time practically {\em all}
spectroscopic indicators are reproduced in a {\em quantitative} manner.  
Consistent results are obtained from lines of the different spin systems,
simultaneously for the neutral {\em and} singly-ionized stages 
(the model atom of Becker \& Butler~(1988) is adopted for O\,{\sc ii}). 
A comparison of non-LTE and LTE abundances for individual lines in a
sub-sample of the test objects is made in Fig.~1. In the main-sequence star
Vega departures from LTE hardly affect the weaker C\,{\sc i} and O\,{\sc i}
lines, whereas the strong lines of these species and the N\,{\sc i} lines
are subject to significant non-LTE effects. These strengthen in the
supergiants, giving rise to non-LTE abundance corrections of typically
$\sim$0.3\,dex for the weak lines and of more than 1\,dex for the strong lines.
By accounting for non-LTE effects systematic trends of derived line
abundance with equivalent width are removed and the statistical scatter from
the weak lines is slightly reduced when compared to LTE analyses. 
In particular, the new models indicate much higher nitrogen abundances from
the analysis of N\,{\sc i} lines in
BA-type supergiants than derived previously, by a factor of $\sim$2. 
The refined collisional excitation data (Frost et al.~1998)
allow for a realistic treatment of the processes trying to restore detailed
balance.
It turns out that the non-LTE departures are drastically dampened
in comparison to the case with less elaborate data. These findings have
significant consequences for the interpretation of the deduced abundance
ratios in terms of chemical mixing in the course of massive star evolution, 
see below and Venn \& Przybilla (VP, these proceedings).

\begin{figure}
\plotone{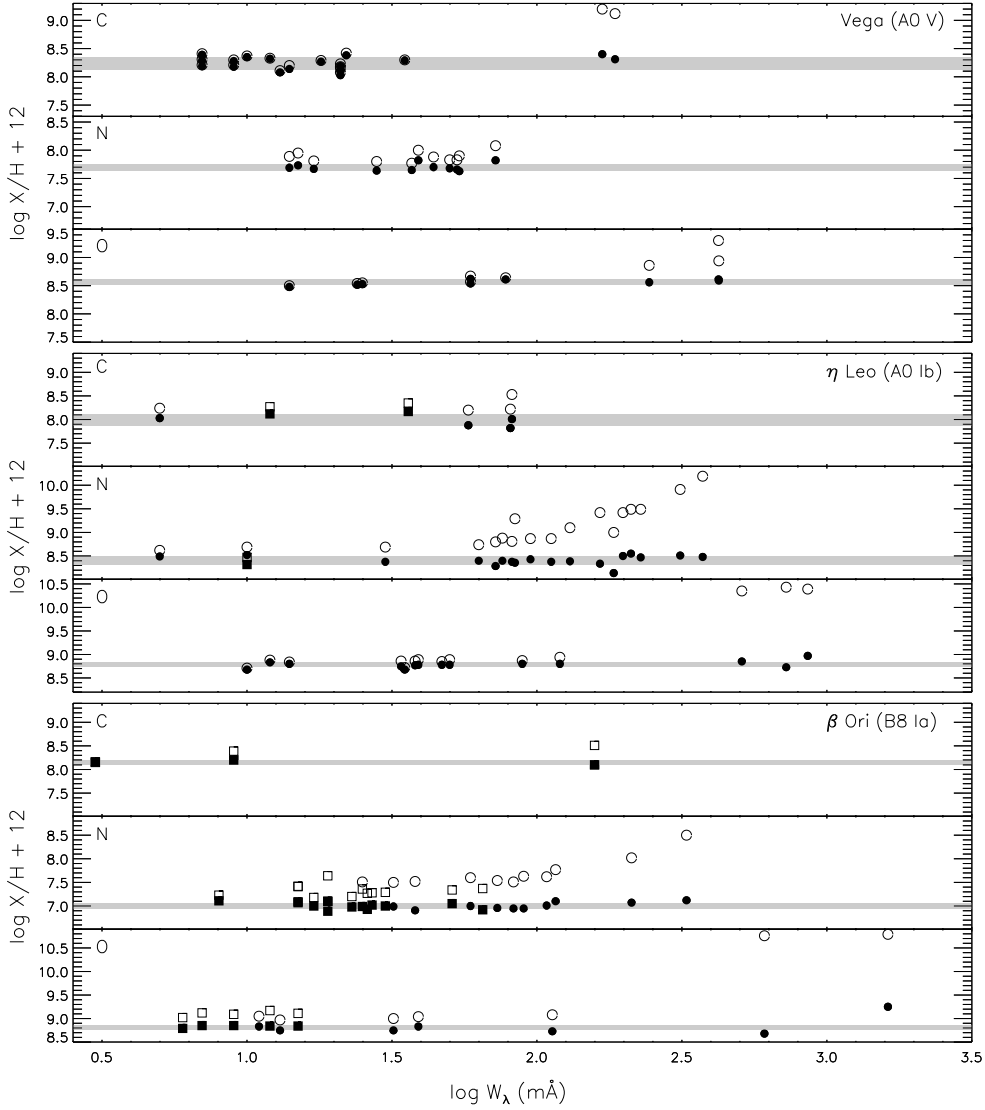}
\caption{Comparison of LTE and non-LTE abundances (open/filled symbols) 
for CNO from our best fits of unblended spectral lines in three bright 
galactic objects
on the usual logarithmic scale. Abundances for the neutral (circles) and 
singly-ionized species (boxes) are displayed as a function of
equivalent width. The grey band spans the uncertainty range associated with
the mean abundances ($\pm1\sigma$ random errors). 
Note that for the derivation of the mean abundances a few additional 
blended lines have also been accounted for by means of spectrum synthesis.}
\end{figure}

To conclude, the new model calculations reduce the random scatter in the
spectral line analyses and largely remove systematic error. They indicate 
that {\em absolute} abundances with 1$\sigma$-uncertainties of
0.05 to 0.10\,dex (random) and $\sim$0.10\,dex (systematic error)
can be derived in main sequence to supergiant stars alike. In addition, the
presence of lines from two ionic species of the elements 
allows to exploit the ionization equilibria for the stellar parameter
determination.

A few issues remain to be solved. The comparatively large uncertainty in the 
carbon abundances is interpreted as an indication for the need of further 
improvement of the atomic data. Discrepant abundances from the C\,{\sc ii}
$\lambda\lambda$\,6578--82 doublet, formed in the wing of H$\alpha$, 
result from the neglect of sphericity and mass-loss effects on the H$\alpha$ 
feature, leading to inappropriate line-formation depths for the C\,{\sc ii} 
lines in the current modelling. Sphericity and mass-loss are also affecting
the oxygen triplet $\lambda\lambda$\,7771--5 and O\,{\sc i}
$\lambda$\,8446, which belong to the strongest lines in luminous
supergiants, sampling the body of the stellar atmosphere. The N\,{\sc i} 
$\lambda\lambda$\,8184--8242 multiplet analysis currently suffers from
problems with realistically modelling the line merging and level 
dissolution of hydrogen near the Paschen series limit, again leading to 
inappropriate line-formation depths for the N\,{\sc i} features. Finally, 
the strongest line in the doublet spin system, N\,{\sc i} $\lambda$\,8629, 
turns out to indicate too low abundances still, despite significant 
improvement compared to previous studies. This line is extremely sensitive 
to inaccuracies in the atomic data and the atmospheric structure.

\section{First applications}
Stellar evolution models accounting for the effects of mass-loss and
rotation have become available recently (Meynet \& Maeder~2000; Heger \& 
Langer 2000). These reproduce many of the observational constraints
(e.g.~A.~Herrero, these pro\-ceedings; VP) 
on massive star evolution at least qualitatively, in particular the surface
contamination with CN-cycle products. In addition
to convective mixing during the red-supergiant stage (first dredge-up) a
second channel opens, rotationally induced mixing. This can account for
abundance anomalies of the light elemental species already on the 
main sequence. 

\begin{figure}[ht!]
\plotone{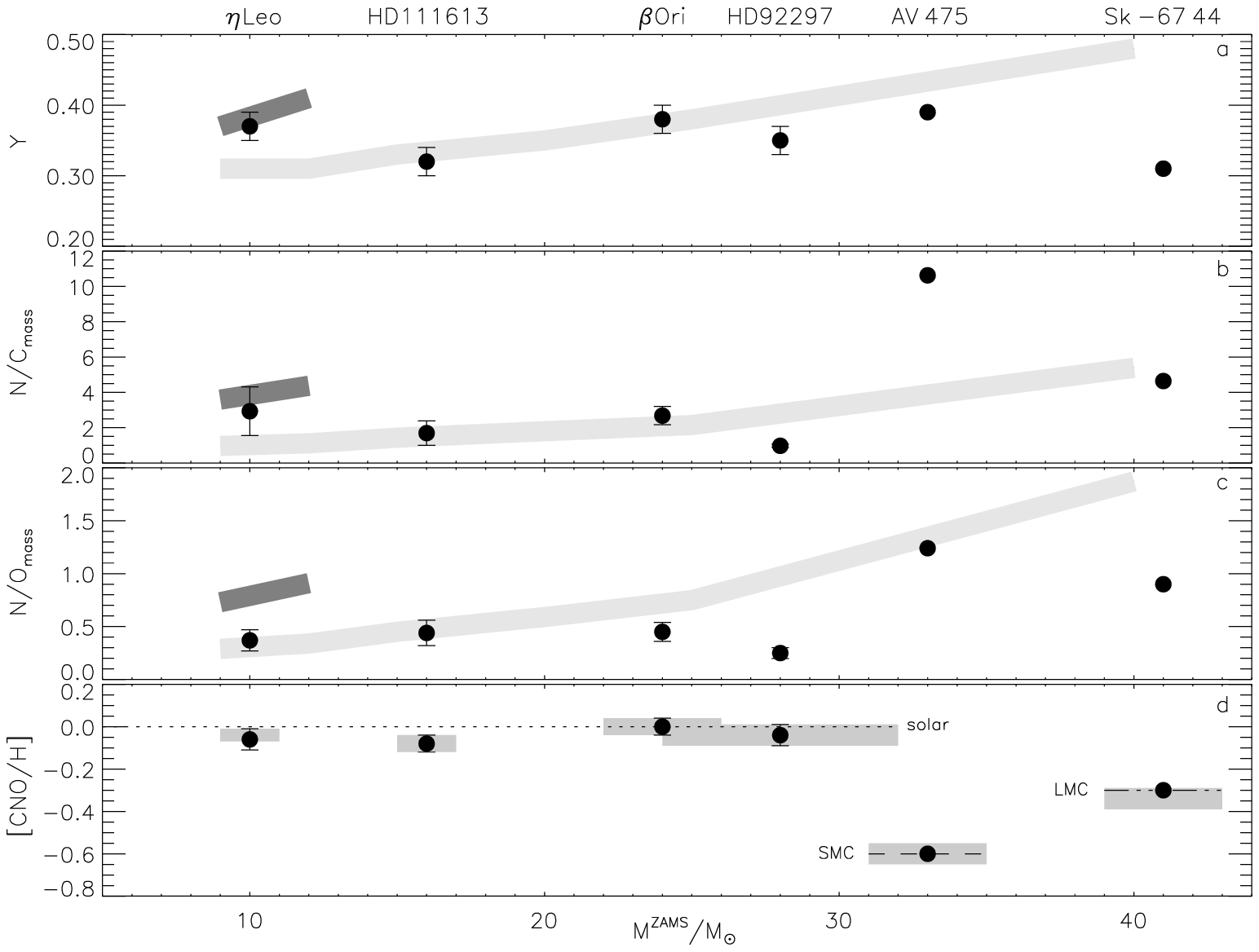}
\caption{Indicators for mixing of nuclear-processed matter into atmospheric
layers: (a) surface He abundance (by mass), (b) N/C and (c) N/O
mass ratio. The light-grey and dark-grey bands denote predictions from 
stellar evolution models (Meynet \& Maeder~2000) for objects 
crossing the HRD for the first time and in the blue-loop phase, respectively
(at $T_{\rm eff}$\,=\,10$^4$\,K, for 
solar metallicity and $v^{\rm ini}_{\rm rot}$\,=\,300\,km\,s$^{-1}$). 
Results from the abundance analyses of
four galactic BSG and the visually brightest stars in both Magellanic Clouds
are indicated by dots. A comparison of the sum of CNO
abundances with the stellar metallicity (as defined by the heavier
elements, grey bands) is made in (d), the values are given relative to 
the solar standard (Grevesse \& Sauval~1998). The width
and the breadth of the bands in (d) indicates the 
$\pm$1$\sigma$ uncertainties in metallicity and ZAMS mass, respectively.
Horizontal lines mark the expected metallicities in the three galactic 
environments.}
\end{figure}

We compare the predictions from the new stellar evolution 
models with the findings from our test sample and preliminary 
results on the visually brightest, i.e. the most massive, A-type supergiants 
in both Magellanic Clouds in Fig.\,2.
Non-LTE abundances for the fusion product helium have been determined in
addition to data on CNO. The objects are assumed to be single stars, as 
binary evolution will alter the interpretation of the results significantly. 
To summarise, good agreement is found in terms of helium enrichment and N/C
ratios. For the least-massive galactic object a blue-loop scenario
with first dredge-up abundances is indicated and the more massive objects have 
most likely evolved directly from the main sequence, allowing for variations 
in the initial rotational velocities.
Far stronger mixing acts in the SMC supergiant, which is
well understood if metallicity effects on stellar evolution are
accounted for, that affect mass-loss and angular-momentum transport
(Maeder \& Meynet~2001). In addition, good agreement between the combined CNO
abundances and stellar metallicity (relative to the solar standard, Grevesse
\& Sauval~1998) is found, as can be expected for the catalysts of the main 
fusion cycle.

However, the consistent picture is shattered when the N/O ratios are
considered. These deviate from the predictions, as oxygen remains practically
undepleted. The reason for this has to be
investigated. Moreover, a CNO overabundance relative to the heavier metals
is indicated, if the solar abundances from 3D hydrodynamical simulations
(M.~Asplund, these proceedings) are preferred over the results from classical
1D model atmospheres.

How do our conclusions compare with those of previous investigations?  
Until recently, CNO abundances in galactic and SMC AF-type supergiants
in the mass-range $\sim$5--20\,M$_{\odot}$ (Venn~1995, 1999) seemed to 
indicate a direct evolution from the main 
sequence, with no necessity to invoke blue-loop scenarios. In view of the
improved non-LTE abundance analysis -- in particular with respect to 
nitrogen -- the situation has to be reinvestigated, with an important 
first step taken by VP. Blue-loop evolution can no longer be ruled out, and
large efforts will be required to disentangle the different alternatives
for explaining the observational findings, see VP for details. An extension
of the sample size is mandatory, including many more objects of masses above
$\sim$20\,M$_{\odot}$, where the signatures of rotationally induced mixing
are supposed to get more pronounced.  

A related field of application is the analysis of {\em unevolved} B-type
main-sequence stars. Here, Korn et al.~(2002) use the new model atoms to
derive pristine stellar CNO abundances in the LMC, based on VLT/UVES 
observations. These compare well with results from H\,{\sc ii} region 
analyses and confirm the extraordinary low present-day nitrogen abundance of
this galactic environment.

\section{Perspective}
The future certainly belongs to extensive applications of the new CNO model 
atoms for quantitative spectroscopy. Efficient multi-object spectrographs 
like FLAMES on the VLT will provide the plethora of observations 
needed to systematically investigate massive star evolution throughout the 
HRD. On the theoretical side, the model atoms will see occasional updates with 
improved atomic data, whenever it becomes available. Furthermore, an
implementation of the model atoms for non-LTE computations of line-blanketed, 
spherically extended stellar atmospheres accounting for mass-loss is aspired, 
in order to solve the remaining few problems with the current approach.
Finally, an extension of the applications towards later-type stars should 
also be considered. This will however require major efforts from theoretical
atomic physics to generate reliable data on hydrogen collisions, which can 
presently be treated in a highly approximative manner only.

\end{document}